\documentclass{article} 
\usepackage{arxiv}
\usepackage{natbib}

\usepackage[pagebackref=true,breaklinks=true,colorlinks,bookmarks=false,citecolor=blue,linkcolor=blue]{hyperref}

\usepackage{url}
\usepackage{booktabs}
\usepackage{multirow}
\usepackage{colortbl}
\usepackage{xcolor}
\usepackage{enumitem}
\usepackage{graphicx}
\usepackage{longtable}
\usepackage{caption}
\usepackage{amssymb}
\usepackage{siunitx}
\usepackage{tabularx}

\definecolor{ConflictColor}{RGB}{255, 220, 220}
\definecolor{CatastrophicColor}{RGB}{153, 0, 0}

\usepackage{booktabs}
\usepackage{multirow}
\usepackage{makecell}
\usepackage[table]{xcolor}
\usepackage{graphicx}
\usepackage{pgf}

\definecolor{ConflictRed}{HTML}{B2182B}

\newcommand{\rrrshadecell}[1]{%
  \pgfmathtruncatemacro{\rrrshade}{%
    min(80,max(5,5+75*min(1,#1)))%
  }%
  \edef\rrrcolor{ConflictRed!\rrrshade!white}%
  \expandafter\cellcolor\expandafter{\rrrcolor}%
}

\newcommand{\rrrconflict}[1]{%
  \rrrshadecell{#1}%
  \textbf{\ensuremath{#1}}%
}

\newcommand{\mypara}[1]{\smallskip\noindent\textbf{#1}}

\title{Defenses at Odds: Measuring and Explaining Defense Conflicts in Large Language Models}

\author{
Xiangtao Meng\textsuperscript{1}\ \ \
Wenyu Chen\textsuperscript{1}\ \ \
Chuanchao Zang\textsuperscript{1}\ \ \
Xinyu Gao\textsuperscript{1}\ \ \
Jianing Wang\textsuperscript{1}\ \ \
\\
Li Wang\textsuperscript{1}\ \ \
Zheng Li\textsuperscript{1}\ \ \
Shanqing Guo\textsuperscript{1}
\\
\\
\textsuperscript{1}\textit{Shandong University} \ \ \ 
}

\begin{document}

\maketitle


\providecommand{\pending}[1]{\textnormal{[#1]}}
\providecommand{\neval}{\pending{N evaluated}}
\providecommand{\nconflict}{\pending{N conflict}}

\begin{abstract}
Large language models (LLMs) may require additional defenses after deployment as risks and governance requirements evolve. 
Subsequent defenses can interact with earlier defenses, raising the question of their sequential compatibility. 
We study this question with \textsc{ConflictEval}, evaluating 144 ordered compositions of six defenses spanning safety, privacy, and fairness across six models. 
The resulting interactions are heterogeneous across defense pairs, application orders, and models. 
Notably, some compositions exhibit defense conflicts: the subsequent defense improves its target objective while weakening protection established by the earlier defense. 
We investigate these interactions through the directional compatibility of defense-induced changes in risk-relevant representations.
Across 11 selected cases, we use activation interventions to assess which defense-induced representational changes support protection and examine how subsequent defenses affect these changes. 
We find that, in some conflict cases, subsequent defenses counteract representational changes supporting earlier protection, providing evidence for one possible pathway to defense conflicts.
Building on this analysis, we propose \emph{Conflict-Triggered Directional Retention} (CTDR), which penalizes opposing shifts along directions supporting earlier protection. On six selected conflicting compositions from this analysis, CTDR reduces first-objective regression while maintaining positive gains on the subsequent defense objective.
\end{abstract}

\section{Introduction}
\label{sec:introduction}
Post-training defenses help large language models (LLMs) meet safety, privacy, and fairness requirements. Approaches such as safety alignment~\citep{rafailov2023direct,ouyang2022training,xhonneux2024efficient}, machine unlearning~\citep{bourtoule2021machine,zhang2024negative,li2024wmdp}, and debiasing~\citep{ilharco2022editing} aim to reduce harmful responses, limit memorized-data leakage, and mitigate stereotypical biases, respectively. However, LLMs may require additional defenses after deployment as new vulnerabilities emerge~\citep{deepmind2024frontier,openai2025preparedness}. For example, a model already aligned for safety may later receive a data-removal request, requiring an update that reduces memorized information while preserving existing protections. Such evolving requirements can lead to defenses being applied sequentially, with each update acting on a model already shaped by earlier defenses. This raises a practical question: \textbf{\textit{when a subsequent defense improves its target objective, does it preserve, strengthen, or weaken the protection established by earlier defenses?}}

A defense's effectiveness on its target risk alone does not establish its compatibility with protections already in place. Prior work~\citep{qi2023fine,bhardwaj2024language,qi2025safety} has shown that subsequent fine-tuning, even on benign tasks, can weaken the safety alignment of LLMs. When the subsequent update is itself a defense, it may improve its target objective while weakening protection established by an earlier defense. We refer to this outcome as a \emph{defense conflict}. Such conflicts highlight the need to evaluate the gains from a subsequent defense alongside the retention of earlier protection.

To examine sequential defense compatibility, we introduce \textsc{ConflictEval} and evaluate 144 ordered compositions of six defenses spanning safety, privacy, and fairness across six models. We jointly evaluate the subsequent defense's target-objective gain and the Relative Regression Rate (RRR), which quantifies how much of the first defense's achieved risk reduction is reversed by the subsequent update. Our evaluation reveals heterogeneous interactions across defense pairs, application orders, and models: subsequent defenses can strengthen earlier protection in some configurations while weakening it in others. Notably, some compositions exhibit defense conflicts in which the subsequent defense improves its own target objective while eroding earlier protection, showing that an effective defense update can come at the expense of protection already established.

We investigate a possible representational explanation for these conflicts: subsequent defenses may weaken earlier protection by counteracting defense-induced changes in risk-relevant representations. Across 11 selected cases, we use activation interventions~\citep{meng2022locating} to assess which defense-induced changes along risk-relevant directions contribute to protection, and examine how subsequent updates affect them. In some conflict cases, later updates oppose these defense-supporting shifts, providing evidence for representational counteraction as one pathway to the observed protection loss. Building on this analysis, we propose \emph{Conflict-Triggered Directional Retention} (CTDR), which penalizes opposing representational shifts along directions supporting earlier protection during subsequent defense training. We evaluate CTDR on six selected conflicting compositions from the mechanism analysis, where it reduces first-objective regression while maintaining positive gains on the subsequent defense objective.

\begin{figure}[t]
  \centering
  \includegraphics[width=0.99\textwidth]{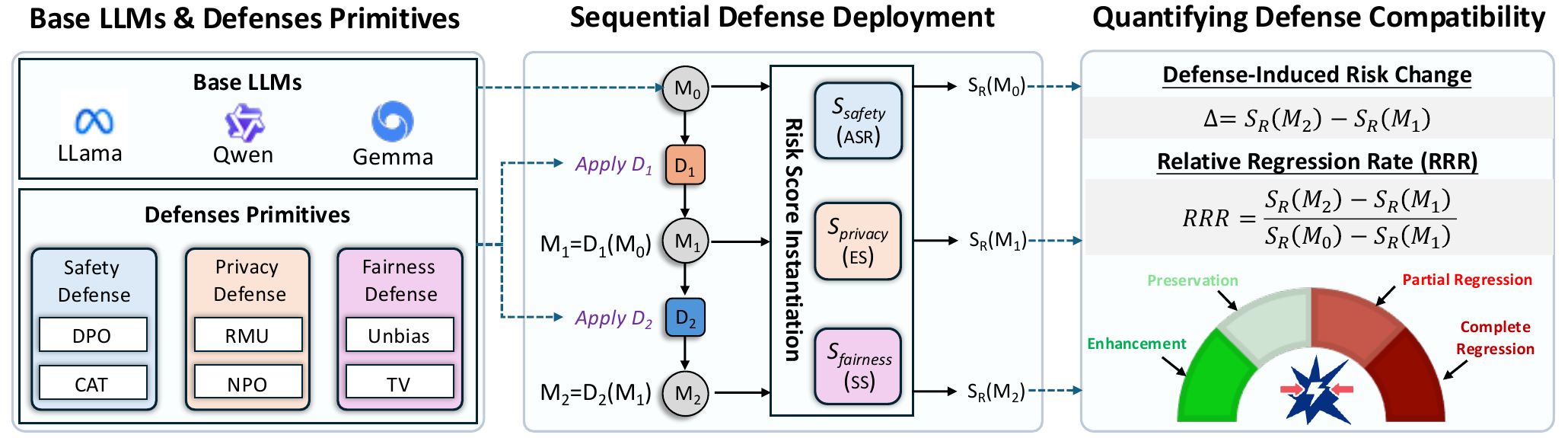} 
  \caption{Overview of the \textsc{ConflictEval} evaluation framework, which quantifies cross-defense interference under sequential deployment ($D_1 \to D_2$) across safety, privacy, and fairness dimensions via the Relative Regression Rate~(RRR).}
  \label{fig:conflicteval}
\end{figure}

\section{Measuring Sequential Defense Compatibility}
\label{sec:evaluation}

\subsection{Evaluation Framework}
\label{subsec:measurement-framework}

To examine sequential defense compatibility, we introduce \textsc{ConflictEval}, an evaluation framework that jointly measures the gains from a subsequent defense and the retention of protection established by an earlier defense. We instantiate the framework with six defenses spanning safety, privacy, and fairness across six models, yielding 144 ordered compositions. As illustrated in ~\autoref{fig:conflicteval}, the framework tracks both target risks across successive model checkpoints and quantifies whether new defense gains accompany the preservation or regression of earlier protection.

\mypara{Sequential composition.}
We consider two configured defense procedures, $D_1$ and $D_2$, applied sequentially. In this study, we focus on cross-risk compositions in which the two defenses target distinct risk dimensions, $R_1 \neq R_2$.
Starting from an initial model $M_0$, the first defense produces $M_1$, and the subsequent defense updates this already defended model to obtain $M_2$:
\begin{equation}
    M_1 = D_1(M_0),
    \qquad
    M_2 = D_2(M_1).
\label{eq:measurement-sequence}
\end{equation}
Here, each defense denotes its configured algorithm, training data, hyperparameters, and update procedure.
Thus, an ordered composition $D_1 \rightarrow D_2$ specifies both the two instantiated defenses and the sequence in which they are applied to the model.

\mypara{Quantifying compatibility.}
Let $S_R(M)$ denote the risk score of model $M$ on dimension $R$, with larger values indicating higher risk.
For each risk dimension, all three checkpoints are evaluated on the
same fixed evaluation set.
To respect this sequential setup, we measure the gain of each defense relative to the checkpoint from which it is applied:
\begin{equation}
\begin{aligned}
G_1 &= S_{R_1}(M_0)-S_{R_1}(M_1), \\
G_2 &= S_{R_2}(M_1)-S_{R_2}(M_2).
\end{aligned}
\label{eq:measurement-gains}
\end{equation}
Positive $G_i$ indicates a reduction in the corresponding target risk.
In particular, $G_2$ is measured from $M_1$ rather than $M_0$, so that it captures the incremental change associated with $D_2$ rather than changes already induced by $D_1$.
The subsequent defense's effect on earlier protection is
\begin{equation}
\Delta_1 = S_{R_1}(M_2)-S_{R_1}(M_1),
\label{eq:measurement-regression}
\end{equation}
where positive $\Delta_1$ indicates regression of earlier protection and negative $\Delta_1$ indicates further risk reduction.
Because the same increase in $R_1$ can reverse different fractions of the risk reduction achieved by $D_1$, we normalize $\Delta_1$ by $G_1$ to obtain the Relative Regression Rate:
\begin{equation}
\mathrm{RRR} = \frac{\Delta_1}{G_1}.
\label{eq:rrr}
\end{equation}
RRR is defined only when $G_1>\tau$, where $\tau=10^{-8}$ is a numerical tolerance. Positive RRR indicates regression of the first-defense gain, with larger values corresponding to greater regression; $\mathrm{RRR}=1$ corresponds to a return to the initial risk level. Negative RRR indicates further reduction of the earlier risk. We identify an observed \emph{defense conflict} when the subsequent
defense reduces its own target risk while protection established by the
earlier defense regresses.


\subsection{Evaluation Setup}
\label{subsec:measurement-protocol}

\mypara{Models and initialization.}
We evaluate six instruction-tuned models from the Llama, Gemma, and Qwen families, selecting a smaller (S) and a larger (L) model from each family (see \autoref{tab:target-llms}). To provide a controlled setting for evaluating privacy defenses, we adopt the TOFU initialization procedure used in unlearning benchmarks \citep{maini2024tofu,dorna2025openunlearning}. TOFU contains question--answer pairs about fictitious authors, allowing models to acquire identifiable target knowledge through fine-tuning before unlearning is applied. We fine-tune each model on this dataset and use the resulting checkpoint as $M_0$. All defense compositions for a given model start from this same checkpoint, enabling comparisons of sequential updates from a common initial state.

\begin{table}[h]
\centering
\caption{
Target LLMs used in \textsc{ConflictEval}. S and L denote the smaller and larger models.
}
\label{tab:target-llms}

\begingroup
\small
\setlength{\tabcolsep}{6pt}
\renewcommand{\arraystretch}{1.12}

\begin{tabular}{@{}llr@{}}
\toprule
Model family & Short name & Model identifier \\
\midrule

\multirow{2}{*}{Llama}
& Llama-S
& Llama-3.2-1B-Instruct~\citep{meta2024llama32_1b_instruct} \\
& Llama-L
& Meta-Llama-3-8B-Instruct~\citep{meta2024llama3_8b_instruct} \\

\addlinespace[4pt]

\multirow{2}{*}{Gemma}
& Gemma-S
& gemma-2-2b-it~\citep{google2024gemma2_2b_it} \\
& Gemma-L
& gemma-7b-it~\citep{google2024gemma_7b_it} \\

\addlinespace[4pt]

\multirow{2}{*}{Qwen}
& Qwen-S
& Qwen2.5-1.5B-Instruct~\citep{qwen2024_qwen25_15b_instruct} \\
& Qwen-L
& Qwen2.5-7B-Instruct~\citep{qwen2024_qwen25_7b_instruct} \\

\bottomrule
\end{tabular}
\endgroup
\end{table}

\begin{table}[t]
\centering
\caption{
Relative Regression Rate (RRR) across all ordered defense
compositions.
For positive RRR, darker red indicates greater regression of the
first-defense gain; color intensity increases linearly from
$\mathrm{RRR}=0$ to $\mathrm{RRR}=1$.
Color intensity increases linearly from RRR = 0 to RRR = 1.
Non-positive values are left unshaded.
}
\label{tab:rrr_all_scenarios}

\begingroup
\scriptsize
\setlength{\tabcolsep}{3.4pt}
\renewcommand{\arraystretch}{1.08}

\resizebox{\linewidth}{!}{%
\begin{tabular}{@{}lllrrrrrr@{}}
\toprule

\multirow{2}{*}{\makecell[c]{Deployment\\Scenario}}
&
\multirow{2}{*}{\makecell[c]{Primary Defense\\($D_1$)}}
&
\multirow{2}{*}{\makecell[c]{Subsequent Defense\\($D_2$)}}
&
\multicolumn{2}{c}{Llama}
&
\multicolumn{2}{c}{Gemma}
&
\multicolumn{2}{c}{Qwen}
\\

\cmidrule(lr){4-5}
\cmidrule(lr){6-7}
\cmidrule(lr){8-9}

&
&
&
-S & -L
&
-S & -L
&
-S & -L
\\

\midrule


\multirow{8}{*}{\textbf{Fairness-First}}

&
\multirow{4}{*}{
  \makecell[l]{
    \textbf{Unbias}\\[-1pt]
    \scriptsize(Fairness Stage)
  }
}
&
+ RMU (Privacy)
&
$-0.005$
&
$-0.152$
&
\rrrconflict{0.154}
&
\rrrconflict{0.215}
&
\rrrconflict{0.027}
&
\rrrconflict{0.072}
\\

&
&
+ NPO (Privacy)
&
$-0.401$
&
$-0.222$
&
\rrrconflict{0.174}
&
$-0.316$
&
\rrrconflict{0.012}
&
\rrrconflict{0.201}
\\

&
&
+ DPO (Safety)
&
$-0.284$
&
$-0.181$
&
\rrrconflict{0.403}
&
$-0.300$
&
\rrrconflict{0.048}
&
$-0.078$
\\

&
&
+ CAT (Safety)
&
\rrrconflict{0.600}
&
\rrrconflict{0.724}
&
\rrrconflict{0.637}
&
\rrrconflict{0.301}
&
\rrrconflict{0.659}
&
\rrrconflict{0.526}
\\

\cmidrule(lr){2-9}

&
\multirow{4}{*}{
  \makecell[l]{
    \textbf{Task Vector (TV)}\\[-1pt]
    \scriptsize(Fairness Stage)
  }
}
&
+ RMU (Privacy)
&
\rrrconflict{0.014}
&
$-0.134$
&
$-0.887$
&
\rrrconflict{0.389}
&
$-0.091$
&
$-0.495$
\\

&
&
+ NPO (Privacy)
&
\rrrconflict{0.011}
&
$-0.169$
&
\rrrconflict{0.311}
&
$-0.0006$
&
\rrrconflict{0.029}
&
\rrrconflict{0.058}
\\

&
&
+ DPO (Safety)
&
\rrrconflict{0.091}
&
$-0.001$
&
$-0.438$
&
\rrrconflict{0.158}
&
\rrrconflict{0.018}
&
$-0.011$
\\

&
&
+ CAT (Safety)
&
\rrrconflict{0.403}
&
\rrrconflict{0.346}
&
\rrrconflict{0.575}
&
\rrrconflict{0.188}
&
\rrrconflict{0.609}
&
\rrrconflict{0.218}
\\

\midrule


\multirow{8}{*}{\textbf{Privacy-First}}

&
\multirow{4}{*}{
  \makecell[l]{
    \textbf{NPO}\\[-1pt]
    \scriptsize(Privacy Stage)
  }
}
&
+ Unbias (Fairness)
&
$-0.002$
&
$-0.564$
&
$0$
&
$0$
&
$-0.078$
&
$-0.0006$
\\

&
&
+ TV (Fairness)
&
\rrrconflict{0.125}
&
$-0.639$
&
$0$
&
$0$
&
$-0.079$
&
$-0.001$
\\

&
&
+ DPO (Safety)
&
$-0.003$
&
$-0.463$
&
$0$
&
$0$
&
$-0.078$
&
\rrrconflict{0.0004}
\\

&
&
+ CAT (Safety)
&
\rrrconflict{0.045}
&
$-0.656$
&
\rrrconflict{0.037}
&
\rrrconflict{0.016}
&
$-0.009$
&
\rrrconflict{0.045}
\\

\cmidrule(lr){2-9}

&
\multirow{4}{*}{
  \makecell[l]{
    \textbf{RMU}\\[-1pt]
    \scriptsize(Privacy Stage)
  }
}
&
+ Unbias (Fairness)
&
$-0.059$
&
$-0.277$
&
$-0.170$
&
$-0.190$
&
$-0.379$
&
$-0.104$
\\

&
&
+ TV (Fairness)
&
$-0.057$
&
$-0.255$
&
$-0.171$
&
$-0.196$
&
$-0.869$
&
$-0.224$
\\

&
&
+ DPO (Safety)
&
$-0.058$
&
$-0.261$
&
\rrrconflict{0.022}
&
$-0.158$
&
\rrrconflict{0.017}
&
$-0.070$
\\

&
&
+ CAT (Safety)
&
\rrrconflict{0.053}
&
$-0.250$
&
$-0.055$
&
$-0.181$
&
$-0.356$
&
$-0.130$
\\

\midrule


\multirow{8}{*}{\textbf{Safety-First}}

&
\multirow{4}{*}{
  \makecell[l]{
    \textbf{DPO}\\[-1pt]
    \scriptsize(Safety Stage)
  }
}
&
+ Unbias (Fairness)
&
$-0.265$
&
$-0.968$
&
\rrrconflict{0.798}
&
\rrrconflict{0.053}
&
$-0.146$
&
$-0.974$
\\

&
&
+ TV (Fairness)
&
$-0.473$
&
$-1.879$
&
\rrrconflict{0.505}
&
$-0.391$
&
$-1.457$
&
$-2.864$
\\

&
&
+ NPO (Privacy)
&
\rrrconflict{0.228}
&
$-0.681$
&
\rrrconflict{0.697}
&
$-0.314$
&
$-0.477$
&
$-0.064$
\\

&
&
+ RMU (Privacy)
&
\rrrconflict{0.310}
&
$-0.315$
&
\rrrconflict{0.132}
&
\rrrconflict{0.008}
&
$-0.132$
&
$-0.206$
\\

\cmidrule(lr){2-9}

&
\multirow{4}{*}{
  \makecell[l]{
    \textbf{CAT}\\[-1pt]
    \scriptsize(Safety Stage)
  }
}
&
+ Unbias (Fairness)
&
$-0.120$
&
$-0.105$
&
$-0.361$
&
\rrrconflict{0.409}
&
\rrrconflict{0.039}
&
\rrrconflict{0.002}
\\

&
&
+ TV (Fairness)
&
$-0.094$
&
\rrrconflict{0.032}
&
$-0.264$
&
$-0.348$
&
\rrrconflict{0.301}
&
\rrrconflict{0.417}
\\

&
&
+ NPO (Privacy)
&
$-0.101$
&
$-0.187$
&
\rrrconflict{0.090}
&
\rrrconflict{0.222}
&
\rrrconflict{0.085}
&
$-0.004$
\\

&
&
+ RMU (Privacy)
&
$-0.173$
&
$-0.011$
&
$-0.097$
&
\rrrconflict{0.038}
&
$-0.069$
&
$-0.003$
\\

\bottomrule
\end{tabular}%
}

\endgroup
\end{table}

\mypara{Defenses and coverage.}
We instantiate two defenses for each risk dimension, chosen to cover distinct update mechanisms. The safety defenses are DPO~\citep{rafailov2023direct}, based on preference optimization over safe and unsafe responses, and CAT~\citep{xhonneux2024efficient}, based on adversarial training with continuous input perturbations. The privacy defenses are NPO~\citep{zhang2024negative}, which applies a negative preference objective to forget data, and RMU~\citep{li2024wmdp}, which modifies representations associated with targeted content. For fairness, we use Unbias~\citep{liu2025mitigating}, which applies targeted unlearning to stereotypical associations, and task-vector editing (TV~\citep{ilharco2022editing,xu2025biasfreebench}), which counteracts a bias-related parameter update.
For each model, we evaluate every ordered pair of defenses targeting different risk dimensions, yielding $6\times4=24$ ordered compositions per model and 144 overall. Unless otherwise stated, sequential compositions apply the configured defense procedures without any additional mechanism designed to preserve the protection established by $D_1$.
See \autoref{subsec:defense-impl} for defense implementation details.

\mypara{Risk Score Instantiation.}
We instantiate $S_R$ separately for safety, privacy, and fairness using benchmark-specific metrics. All risk scores are oriented so that larger values indicate higher risk.

\begin{itemize}[leftmargin=*]
    \item \textbf{Safety Risk.}
    We measure safety risk using the \emph{Attack Success Rate} (ASR) on 404 high-risk prompts sourced from SALAD-Bench~\citep{li2024salad}.
    The prompts were drawn from the Malicious Use and Representation \& Toxicity categories (excluding Unfair Representation) and retained if they elicited a harmful response from at least one of the six $M_0$ checkpoints.
    A response is counted as harmful when its MD-Judge
    v0.2~\footnote{\url{https://huggingface.co/OpenSafetyLab/MD-Judge-v0_2-internlm2_7b}}
    score exceeds 3. We average ASR over three decoding seeds using temperature $0.6$, top-$p$ $0.9$, and a maximum of 256 generated
    tokens. 

    \item \textbf{Privacy Risk.} We instantiate privacy risk using mean extraction strength (ES) on the TOFU forget set~\citep{maini2024tofu}, following OpenUnlearning's teacher-forced exact-suffix matching evaluator~\citep{dorna2025openunlearning}. ES measures matching answer suffixes between tokenwise predictions and the target answer, with higher values indicating greater residual target-content predictability. 

    \item \textbf{Fairness Risk.} 
    We measure fairness risk using StereoSet intrasentence examples \citep{nadeem2021stereoset}. Let $\mathrm{SS}(M)$ denote the stereotype preference score, expressed as a percentage. We define $S_{\mathrm{fairness}}(M)=|\mathrm{SS}(M)-50|/50$, the normalized deviation from the balanced point $\mathrm{SS}=50$, such that larger values indicate stronger preference in either direction.
\end{itemize}

These benchmark-specific risk scores form the basis for subsequent compatibility measurements. Defense gains and RRR values are computed from unrounded checkpoint scores.

\subsection{Observed Sequential Compatibility}
\label{subsec:measurement-results}

\autoref{tab:rrr_all_scenarios} summarizes the observed compatibility of the 144 ordered compositions. Full risk scores at each checkpoint are reported in~\autoref{app:risk-scores}.
We additionally evaluate MMLU~\citep{hendrycks2021measuringmassivemultitasklanguage} as a utility check; most compositions remain close to their undefended baselines, suggesting that the observed risk regressions are not attributable to broad capability collapse (~\autoref{app:mmlu}).

\mypara{Defense conflicts are observed across the evaluated compositions.}
Across the 144 ordered compositions, 57 meet our definition of a defense conflict: the subsequent defense reduces its target risk while the risk addressed by the earlier defense increases. Observed conflicts span all three deployment scenarios, occurring in 30 of 48 fairness-first, 18 of 48 safety-first, and 9 of 48 privacy-first compositions. Because each configured composition is represented by a single training realization, these counts are descriptive of the observed runs rather than estimates of how frequently the same configurations would exhibit conflicts across random seeds. We therefore use them to characterize the breadth of observed defense conflicts, not their seed-level prevalence.

\mypara{Observed compatibility varies across defense pairs and models.} 
The observed compatibility patterns are not determined solely by the pair of risk dimensions being defended. For example, on Llama-S, applying NPO after DPO increases safety risk from $0.4670$ to $0.5095$ ($\mathrm{RRR}=0.228$), whereas applying NPO after CAT decreases risk from $0.2954$ to $0.2591$ ($\mathrm{RRR}=-0.101$). 
In both cases, NPO reduces privacy risk, but its effect on the protection established by the preceding safety defense differs. The same ordered composition can also produce different outcomes across models. When NPO is applied after DPO, safety protection regresses on Gemma-S ($\mathrm{RRR}=0.697$), whereas on Gemma-L the same composition further reduces safety risk from $0.2673$ to $0.1444$ ($\mathrm{RRR}=-0.314$), while reducing privacy risk in both cases. These observations show that sequential compatibility cannot be inferred from the targeted risk dimensions alone; within our evaluation, it varies across defense pairs and models.

\mypara{Observed compatibility varies with application order.} The same two defenses can exhibit different compatibility when their order is reversed. On Llama-S, applying RMU after DPO causes regression of the preceding safety protection ($\mathrm{RRR}=0.310$), whereas applying DPO after RMU further reduces the privacy risk established by RMU ($\mathrm{RRR}=-0.058$). Thus, compatibility need not be symmetric: the behavior of one ordering does not determine the reverse. This motivates evaluating defense compositions as ordered sequences rather than unordered pairs.

Taken together, these measurements show that defense conflicts can arise under sequential defense deployment and interactions are heterogeneous across defense pairs, application orders, and models.


\section{Representational Analysis of Defense Conflicts}
\label{sec:mechanism}

The behavioral results motivate a representational hypothesis: subsequent defenses may weaken earlier protection by counteracting representational changes that support it. We investigate this hypothesis by identifying risk-relevant representations, assessing the contribution of defense-induced changes along these representations, and measuring how subsequent updates affect those changes. 

\subsection{Analysis Setup}

\mypara{Case selection.}
We analyze 11 selected cases from the sequential-composition evaluation
(\autoref{tab:mechanism-cases}). These cases include compositions in which subsequent defenses weaken, preserve, or strengthen earlier protection, allowing us to examine representational changes across these different outcomes. They span three model families, both model scales, and all three risk dimensions. Two reverse-order pairs allow us to compare how the same defenses interact under different application orders.

\mypara{Analysis data.}
We use examples from the evaluation datasets described in \autoref{subsec:measurement-protocol} to construct the risk-specific contrasts defined below. We partition the analysis data into discovery, development, and test sets: discovery data determine candidate risk-relevant representations and intervention scales, development data select directions and defense-support locations, and test data evaluate these fixed selections and the representational changes under sequential composition.


\subsection{Identifying Risk-Relevant Representations}
\label{subsec:risk-directions}

We characterize risk-relevant representations as directions in activation spaces. To make intervention evaluation computationally efficient, we use risk proxy scores based on risk-token log probabilities, which capture model preferences for risk-relevant text without full response generation.

\mypara{Estimating candidate risk directions.}
Let $h_\ell(M,x)$ denote the residual activation of input $x$ at Transformer block $\ell$.
For safety, we use the activation at the final prompt token.
For privacy and fairness, we average activations over the selected predictive positions following divergence between paired candidates.

For each risk dimension $R$, we construct two discovery-set contrast groups, $\mathcal{D}_R^A$ and $\mathcal{D}_R^B$.
The contrasts are harmful versus harmless requests for safety, original versus question-matched perturbed answers for privacy,
and stereotypical versus anti-stereotypical sentences for fairness.
We estimate the candidate direction as
\begin{equation}
d_{R,\ell}
=
\frac{1}{|\mathcal{D}_R^A|}
\sum_{x \in \mathcal{D}_R^A} h_\ell(M_0,x)
-
\frac{1}{|\mathcal{D}_R^B|}
\sum_{x \in \mathcal{D}_R^B} h_\ell(M_0,x).
\label{eq:candidate-direction}
\end{equation}
We normalize $d_{R,\ell}$ to obtain a unit vector $u_{R,\ell}$, which serves as the risk direction at layer $\ell$.

\mypara{Evaluating candidate risk directions.}
We assess whether candidate directions influence risk-related model behavior by increasing or decreasing the activation component along each direction. 
For each $u_{R,\ell}$, we intervene on $M_0$ at the analysis positions used for direction estimation:
\begin{equation}
h_\ell \leftarrow h_\ell \pm \alpha \sigma_{R,\ell}u_{R,\ell},
\label{eq:direction-intervention}
\end{equation}
where $\alpha=0.25$ and $\sigma_{R,\ell}$ is the standard deviation across discovery examples of the target-sequence mean projections onto $u_{R,\ell}$. Let $M_0^{+,\ell}$ and $M_0^{-,\ell}$ denote the models under the positive and negative interventions, respectively. We quantify the directional effect as
\begin{equation}
E_{R,\ell}=\bar q_R(M_0^{+,\ell})-\bar q_R(M_0^{-,\ell}),
\label{eq:direction-test}
\end{equation}
where $q_R$ is a risk proxy score derived from token log probabilities. We use this proxy to reduce the cost of repeated intervention evaluations and capture fine-grained changes in model preferences.
For safety risk, it measures the difference in mean token log probability between compliance and refusal prefixes. For privacy risk, it measures the mean token log probability of the original answer. For fairness risk, it measures the stereotype--anti-stereotype log-probability margin.
We select directions with resolved orientation and positive effects above numerical tolerance on development data.


\subsection{Assessing Defense-Supporting Changes}
\label{subsec:functional-support}

We next assess whether defense-induced changes along the identified directions contribute to protection. We test this by restoring their pre-defense projections and measuring whether defense performance deteriorates.

\mypara{Testing defense support.}
For defense $D_i$ targeting $R_i$, we examine its update $M_{i-1}\rightarrow M_i$ using the risk direction $u_{i,\ell}=u_{R_i,\ell}$ identified on $M_0$. 
At each layer with a direction selected in \autoref{subsec:risk-directions}, we restore the activation projection along that direction to its pre-defense value:
\begin{equation}
h_\ell(M_i,x)\leftarrow h_\ell(M_i,x)-u_{i,\ell}u_{i,\ell}^{\top}\bigl[h_\ell(M_i,x)-h_\ell(M_{i-1},x)\bigr].
\label{eq:support-restoration}
\end{equation}
This intervention removes the projected change while preserving orthogonal components at the intervention site. Let $M_i^{\mathrm{restore},\ell}$ denote the model under this intervention. We measure its effect as
\begin{equation}
A_{i,\ell}=\bar q_{R_i}(M_i^{\mathrm{restore},\ell})-\bar q_{R_i}(M_i).
\label{eq:restoration-effect}
\end{equation}
A positive $A_{i,\ell}$ indicates that restoration increases the risk proxy score, suggesting that the defense relies in part on changes along this risk representation.

\mypara{Results.}
We record the layers containing risk-relevant representational changes that support defense $D_i$ on development data as $K_i$. As shown in \autoref{tab:mechanism-cases}, $K_1$ and $K_2$ overlap in 10 of the 11 compositions, sharing between one and eleven layers. Qwen-L Unbias$\rightarrow$CAT is the only case with disjoint support sets. Shared layers place the defense-supporting changes in the same activation space, although their directions may differ. We next examine how subsequent updates affect the changes supporting earlier protection.

\begin{table}[t]
\centering
\caption{Defense-support locations and sequential effects in 11 compositions. $K_1$ and $K_2$ are selected on development data for $D_1(M_0)$ and $D_2(M_1)$, respectively; entries are one-based block indices. RDC is pooled over $K_1\cap K_2$, except $\dagger$, which uses $K_1$. Negative RDC indicates opposition to the first-defense shift; positive RRR indicates earlier-protection regression.}
\label{tab:mechanism-cases}
\begingroup
\small
\setlength{\tabcolsep}{3pt}
\renewcommand{\arraystretch}{1.18}
\begin{tabularx}{\linewidth}{@{}ll
>{\hsize=1.15\hsize\linewidth=\hsize\raggedright\arraybackslash}X
>{\hsize=1.15\hsize\linewidth=\hsize\raggedright\arraybackslash}X
>{\hsize=0.7\hsize\linewidth=\hsize\raggedright\arraybackslash}X
rr@{}}
\toprule
& & \multicolumn{3}{c}{Defense-support locations}
& \multicolumn{2}{c}{Sequential effects} \\
\cmidrule(lr){3-5}
\cmidrule(l){6-7}
Model & Defense sequence & $K_1$ & $K_2$ & $K_1\cap K_2$ & Pooled RDC & RRR \\
\midrule
\textbf{Llama-S}
& DPO$\rightarrow$RMU
& 8--10, 12--15
& 2, 11--13
& 12, 13
& $-1.687$ & $0.310$ \\
& RMU$\rightarrow$DPO
& 2, 11--13
& 8--10, 12--15
& 12, 13
& $0.389$ & $-0.058$ \\
& DPO$\rightarrow$NPO
& 8--10, 12--15
& 4, 8, 11--13
& 8, 12, 13
& $-1.789$ & $0.228$ \\
\addlinespace[5pt]
\textbf{Gemma-S}
& DPO$\rightarrow$NPO
& 1, 2, 7, 15
& 11, 15, 17--26
& 15
& $-0.050$ & $0.697$ \\
& NPO$\rightarrow$DPO
& 11, 15, 17--26
& 1, 2, 7, 15
& 15
& $0.169$ & $0.000$ \\
\addlinespace[5pt]
\textbf{Gemma-L}
& CAT$\rightarrow$NPO
& 10, 11, 13, 21--28
& 16, 23
& 23
& $-0.099$ & $0.222$ \\
& CAT$\rightarrow$TV
& 10, 11, 13, 21--28
& 5, 12, 24
& 24
& $1.206$ & $-0.348$ \\
\addlinespace[5pt]
\textbf{Qwen-S}
& Unbias$\rightarrow$DPO
& 1, 2, 4, 6, 7, 14, 17, 19--27
& 4, 23--28
& 4, 23--27
& $-0.054$ & $0.048$ \\
& TV$\rightarrow$NPO
& 1--3, 5, 6, 14--21, 23, 27
& 2, 9, 13--27
& 2, 14--21, 23, 27
& $-0.014$ & $0.029$ \\
\addlinespace[5pt]
\textbf{Qwen-L}
& Unbias$\rightarrow$CAT
& 5--10
& 2, 3, 12, 13, 16, 20, 24, 25, 28
& $\varnothing$
& $-2.207$\rlap{$^{\dagger}$} & $0.526$ \\
& NPO$\rightarrow$CAT
& 11, 12, 15, 16, 21, 25
& 2, 3, 12, 13, 16, 20, 24, 25, 28
& 12, 16, 25
& $-1.322$ & $0.045$ \\
\bottomrule
\end{tabularx}
\endgroup
\end{table}


\subsection{Relating Sequential Changes to Regression}
\label{subsec:directional-compatibility}

We next examine whether subsequent updates reinforce or counteract representational changes supporting earlier protection. We measure their directional relationship along the first-risk direction and compare it with behavioral regression.

\mypara{Measuring directional compatibility.}
For each input $x$, we project both sequential updates onto the fixed first-risk direction $u_{1,\ell}$:
\begin{align}
a_\ell(x) &= u_{1,\ell}^{\top}\bigl[h_\ell(M_1,x)-h_\ell(M_0,x)\bigr],
\label{eq:first-defense-projection}\\
b_\ell(x) &= u_{1,\ell}^{\top}\bigl[h_\ell(M_2,x)-h_\ell(M_1,x)\bigr].
\label{eq:sequential-projections}
\end{align}
Here, $a_\ell(x)$ captures the first-defense change, while $b_\ell(x)$ captures the subsequent change along the same direction. Their product is negative when the two updates oppose each other. We compute these projections on test examples for the first risk, using the same activation positions as in the preceding analysis.

Let $K=K_1\cap K_2$ denote the shared support layers and $\mathcal{T}_{R_1}$ the first-risk test set. We define pooled Representational Directional Compatibility (RDC) as
\begin{equation}
\mathrm{RDC}(K)
=
\frac{
\sum_{\ell\in K}\sum_{x\in\mathcal{T}_{R_1}}
a_\ell(x)b_\ell(x)
}{
\sum_{\ell\in K}\sum_{x\in\mathcal{T}_{R_1}}
a_\ell(x)^2
}.
\label{eq:pooled-rdc}
\end{equation}
The denominator normalizes the interaction by the squared magnitude of the first-defense projections. Negative RDC indicates aggregate opposition to the first-defense changes, whereas positive RDC indicates alignment. RDC is an uncentered regression coefficient and can exceed one in magnitude. Its magnitude does not directly quantify behavioral regression. For Qwen-L Unbias$\rightarrow$CAT, the shared set is empty. We therefore compute RDC over $K_1$ to examine how CAT affects the changes supporting Unbias, and mark this result separately in \autoref{tab:mechanism-cases}.

\mypara{Results.}
Across the selected compositions, representational counteraction closely tracks behavioral regression. All eight cases in which the earlier protection regresses have negative RDC, meaning that the subsequent defense moves the model against the representational changes induced by the first defense. By contrast, the three cases without regression have positive RDC, indicating that the second update largely preserves or reinforces those earlier changes. This pattern is particularly clear in the reverse-order pairs. For Llama-S, RMU counteracts the changes established by DPO and the earlier protection regresses; reversing the order instead yields aligned changes with no regression. Gemma-S shows the same qualitative pattern for DPO and NPO: the order with opposing changes exhibits regression, whereas the reverse order shows aligned changes and no regression. These cases suggest that the interaction depends not simply on which defenses are combined, but on whether the later update reinforces or counteracts the changes established by the earlier defense.

Qwen-L Unbias$\rightarrow$CAT illustrates a possible interaction outside shared support layers. Although the two support sets are disjoint, CAT produces negative RDC ($-2.207$) at the Unbias support layers, alongside behavioral regression. A defense's support set identifies changes contributing to its own effect, but does not delimit all representations it may modify. This case therefore suggests that a subsequent defense may disrupt earlier protection through changes outside its own identified support layers. Together, these observations support representational counteraction as one possible pathway to defense conflicts and motivate preserving first-defense changes during subsequent training.

\begin{table}[t]
\centering
\caption{CTDR results on six defense conflicts. Sequential denotes composition without CTDR. Retained gives the percentage of sequential $G_2$ preserved by CTDR. Lower RRR indicates better retention of earlier protection; higher $G_2$ indicates greater subsequent-defense gain. Red cells mark reference conflicts, and gray cells show CTDR results. $\dagger$ No shared support layers ($K_1\cap K_2=\varnothing$).}
\label{tab:ctdr}
\begingroup
\small
\setlength{\tabcolsep}{5pt}
\renewcommand{\arraystretch}{1.12}
\begin{tabular}{@{}llrrrr@{\hspace{4pt}}r@{}}
\toprule
\multirow{2}{*}{Model}
& \multirow{2}{*}{Defense sequence}
& \multicolumn{2}{c}{RRR $\downarrow$}
& \multicolumn{3}{c}{$G_2$ $\uparrow$} \\
\cmidrule(lr){3-4}
\cmidrule(lr){5-7}
& & Sequential & CTDR & Sequential & CTDR & Retained \\
\midrule
\multirow{2}{*}{\textbf{Llama-S}}
& DPO$\rightarrow$RMU
& \rrrconflict{0.310}
& \cellcolor{black!6}$-0.053\phantom{0}$
& $0.636$
& \cellcolor{black!6}$0.633$
& $99.5\%$ \\
& DPO$\rightarrow$NPO
& \rrrconflict{0.228}
& \cellcolor{black!6}$-0.434\phantom{0}$
& $0.645$
& \cellcolor{black!6}$0.631$
& $97.8\%$ \\
\addlinespace[3pt]
\textbf{Gemma-S}
& DPO$\rightarrow$NPO
& \rrrconflict{0.697}
& \cellcolor{black!6}$0.000$
& $0.925$
& \cellcolor{black!6}$0.896$
& $96.9\%$ \\
\textbf{Gemma-L}
& CAT$\rightarrow$NPO
& \rrrconflict{0.222}
& \cellcolor{black!6}$-0.439\phantom{0}$
& $0.112$
& \cellcolor{black!6}$0.112$
& $100.0\%$ \\
\addlinespace[3pt]
\multirow{2}{*}{\textbf{Qwen-L}}
& NPO$\rightarrow$CAT
& \rrrconflict{0.045}
& \cellcolor{black!6}$0.000$
& $0.638$
& \cellcolor{black!6}$0.641$
& $100.5\%$ \\
& Unbias$\rightarrow$CAT$^\dagger$
& \rrrconflict{0.526}
& \cellcolor{black!6}$-0.077\phantom{0}$
& $0.687$
& \cellcolor{black!6}$0.693$
& $100.9\%$ \\
\bottomrule
\end{tabular}
\endgroup
\end{table}

\section{Mitigating Defense Conflicts}
\label{sec:mitigation}

The preceding analysis suggests that subsequent defenses may weaken earlier protection by counteracting changes that support it. We therefore test whether preserving these changes can reduce regression while retaining the benefits of the subsequent defense. We propose \emph{Conflict-Triggered Directional Retention} (CTDR), which penalizes opposing changes along the identified first-risk directions during subsequent defense training.

Let $M_\theta$ be initialized from $M_1$ and trained on the second defense objective. We use fixed first-risk development examples as anchors. For each anchor $x$ and support layer $\ell\in K_1$, we retain the first-defense projection $a_\ell(x)$ defined in \autoref{eq:first-defense-projection} and measure the current change as
\begin{equation}
b_\ell(x;\theta)
=
u_{1,\ell}^{\top}\bigl[h_\ell(M_\theta,x)-h_\ell(M_1,x)\bigr].
\label{eq:ctdr-current-shift}
\end{equation}
A negative product $a_\ell(x)b_\ell(x;\theta)$ indicates that the current change opposes the first-defense shift. For an anchor minibatch $\mathcal{B}$, we optimize
\begin{equation}
\mathcal{L}(\theta)
=
\mathcal{L}_{D_2}(\theta)
+
\lambda
\sum_{\ell\in K_1}w_\ell
\frac{
\sum_{x\in\mathcal{B}}
\bigl[-a_\ell(x)b_\ell(x;\theta)\bigr]_+^2
}{
\sum_{x\in\mathcal{B}}a_\ell(x)^2+\epsilon
},
\label{eq:ctdr}
\end{equation}
where $[z]_+=\max(z,0)$, $\mathcal{L}_{D_2}$ is the subsequent defense's training objective, and $\lambda$ controls the retention penalty. We set $\epsilon=10^{-8}$ and weight each layer by its positive development restoration effect:
\begin{equation}
w_\ell
=
\frac{A_{1,\ell}}{\sum_{j\in K_1}A_{1,j}}.
\label{eq:ctdr-weights}
\end{equation}
Thus, layers with larger restoration effects receive greater weight. The penalty activates only for opposing changes and does not directly constrain aligned changes or orthogonal activation components. We apply it over $K_1$, including layers outside the shared support set. 
Reference projections from $M_0$ and $M_1$ are cached.

\mypara{Evaluation setup.}
We evaluate CTDR on the five stronger conflicts in the mechanism analysis (RRR $0.222$--$0.697$) and one weaker conflict observed when CAT follows NPO on Qwen-L (RRR $0.045$), covering both substantial and mild regression. Following \autoref{subsec:measurement-protocol}, we compare CTDR with the reference sequential compositions using RRR for earlier-protection retention and $G_2$ for subsequent-defense gains. We use $\lambda=1.0$ for all six cases and keep the other defense-training settings specified in~\autoref{subsec:defense-impl}.

\mypara{Results.}
CTDR improves the retention of earlier protection while preserving the benefits of subsequent defenses (\autoref{tab:ctdr}). Across the six cases, RRR becomes non-positive or approaches zero, while approximately $97\%$ or more of the subsequent-defense gain is retained. The improvement extends across both the stronger conflicts and the weaker conflict observed when CAT follows NPO on Qwen-L. The retained gains indicate that the constraint still allows the subsequent defense to make progress on its target objective. In these evaluated settings, much of the earlier-objective regression can therefore be mitigated with only a small reduction, if any, in subsequent-defense effectiveness. Moreover, CTDR also mitigates regression when CAT follows Unbias on Qwen-L, despite their disjoint support sets. RRR decreases from $0.526$ to $-0.077$, with no decrease in the reported subsequent-defense gain. This regression may involve incidental changes to general-purpose representations: large updates or optimization noise during the second defense could perturb representations that also support earlier protection. The mitigation result is consistent with CTDR limiting such interference by constraining opposing changes at the first-defense support layers.

\section{Related Work} \label{sec:related} 

\mypara{Sequential updates and defense retention.} Prior work has shown that later model updates can degrade behaviors established by earlier training. Downstream fine-tuning can weaken safety alignment \citep{qi2023fine,bhardwaj2024language}, while machine unlearning can be partially reversed by subsequent fine-tuning, motivating methods that improve post-unlearning robustness~\citep{wang2025invariance}. These studies establish that post-hoc optimization can interfere with earlier protections. We study a distinct setting in which the subsequent update is itself an independently motivated defense targeting a different risk dimension, and systematically evaluate such ordered compositions across safety, privacy, and fairness. 


\mypara{Representational analysis of alignment.} Prior interpretability work has identified localized and low-dimensional representations associated with aligned behavior, including safety-critical layers~\citep{li2024safety}, safety-related activation directions \citep{pan2025hidden}, and readable activation changes induced by fine-tuning \citep{minder2025narrow}. Activation interventions further provide tools for testing whether internal representations causally support model behavior \citep{meng2022locating}. Building on these ideas, we study not an individual defense in isolation, but how representational changes induced by a subsequent defense interact with changes supporting earlier protection, and use this analysis to motivate directional retention during sequential defense training.

\section{Discussion and Conclusion} 
\label{sec:discussion-conclusion} 
\textsc{ConflictEval} studies sequential defense compatibility across safety, privacy, and fairness. 
Across 144 ordered compositions, we find that subsequent defenses can preserve, strengthen, or weaken earlier protection, and that these interactions vary across defense pairs, application orders, and models. In particular, we observe defense conflicts in which a new defense improves its own objective while causing previously established protection to regress, showing that successful defense updates do not necessarily compose safely.
Our representational analysis provides evidence that counteraction of defense-supporting changes is one possible pathway to such conflicts, linking behavioral regression to how successive defenses reshape risk-relevant representations. Building on this analysis, CTDR reduces earlier-objective regression while retaining subsequent-defense gains in six evaluated conflicts. Overall, our results highlight that sequential defenses should be evaluated not only by the gains they introduce, but also by how well they preserve protection already established.

\mypara{Limitations.} Our study has several limitations. First, the 144 compositions are evaluated with a single defense-training realization, so the observed conflicts should not be interpreted as stable properties of particular defense pairs across random seeds. This limitation does not affect our primary goal of establishing that defense conflicts can arise under sequential deployment and that their occurrence depends on defense order and model choice. Second, our representational analysis covers 11 selected cases and therefore does not establish representational counteraction as a universal mechanism for all conflicts. However, these cases span three model families, both model scales, all three risk dimensions, and different interaction outcomes, providing evidence that the observed pattern extends beyond a single model or defense setting. Finally, CTDR is evaluated on six conflicts, so its generality remains to be established; nevertheless, these results serve as a proof of concept that preserving defense-supporting representational changes can mitigate earlier-objective regression while retaining subsequent-defense gains.

\bibliography{iclr2027_conference}
\bibliographystyle{iclr2027_conference}

\appendix
\section{Defense Implementations}
\label{subsec:defense-impl}

We instantiate six defense methods, with two targeting
each of safety, privacy, and fairness.
These methods cover different update mechanisms, including
preference optimization, adversarial training,
representation-based unlearning, and parameter editing.
Below, we describe their objectives and the datasets used
in our evaluation.

\mypara{Safety defenses.}
We evaluate preference-based alignment and continuous
adversarial training.

\begin{itemize}
    \item \textbf{DPO (Direct Preference Optimization).}
    DPO~\citep{rafailov2023direct} optimizes the relative
    likelihood of preferred and dispreferred responses
    using a reference model.
    For safety alignment, preferred responses follow
    safety requirements, while dispreferred responses
    contain unsafe content.
    We instantiate DPO using preference pairs from
    PKU-SafeRLHF~\citep{ji2025pku}.

    \item \textbf{CAT (Continuous Adversarial Training).}
    CAT~\citep{xhonneux2024efficient} trains models against
    adversarial perturbations in the input embedding space.
    It constructs continuous perturbations that challenge
    the model's safety behavior and updates the model
    to maintain safe responses under these perturbations.
    We instantiate CAT using harmful prompts from
    HarmBench~\citep{mazeika2024harmbench}.
\end{itemize}

\mypara{Privacy defenses.}
We evaluate two unlearning methods on the TOFU forget
set~\citep{maini2024tofu}, targeting the influence of
information introduced during initialization.

\begin{itemize}
    \item \textbf{RMU (Representation Misdirection for
    Unlearning).}
    RMU~\citep{li2024wmdp} changes hidden representations
    associated with content to be forgotten.
    Its objective steers activations on forget examples
    toward a fixed random target while encouraging
    activations on retain examples to remain close to
    those of a frozen reference model.
    We apply this representation-based objective using
    TOFU forget and retain data.

    \item \textbf{NPO (Negative Preference Optimization).}
    NPO~\citep{zhang2024negative} treats forget examples
    as dispreferred outputs and reduces their likelihood
    relative to a fixed reference model.
    This preference-based objective discourages the
    reproduction of targeted content.
    We apply NPO to the designated TOFU forget set.
\end{itemize}

\mypara{Fairness defenses.}
We evaluate unlearning-based debiasing and task-vector
editing using StereoSet-based
data~\citep{nadeem2021stereoset}.

\begin{itemize}
    \item \textbf{Unbias.}
    Unbias~\citep{liu2025mitigating} applies targeted
    unlearning to stereotypical associations while
    retaining anti-stereotypical information.
    Its training objective combines these complementary
    components to reduce stereotypical preferences.
    We instantiate the method using StereoSet examples
    containing stereotypical and anti-stereotypical
    associations.

    \item \textbf{TV (Task-Vector Editing).}
    Task-vector editing
    \citep{ilharco2022editing,xu2025biasfreebench}
    modifies model parameters using a task-specific
    parameter difference.
    In our debiasing setting, a bias-related task vector
    is constructed from the difference between a
    bias-amplified model and its reference checkpoint.
    A scaled version of this vector is subtracted from
    the model being edited to counteract the associated
    biased behavior.
    The bias-related training data are derived from
    StereoSet.
\end{itemize}

\mypara{Implementation details.} \autoref{tab:defense-training-config} summarizes the main training configurations used for the six defenses. Together with the objectives and datasets described above, these settings specify the core implementation of each defense in our evaluation.

\begin{table}[t]
\centering
\caption{
Primary training configurations for the six defense methods.
Batch size follows the convention of the corresponding implementation.
}
\label{tab:defense-training-config}

\begingroup
\small
\setlength{\tabcolsep}{11pt}
\renewcommand{\arraystretch}{1.08}

\begin{tabular}{@{}lccc@{}}
\toprule
Defense & Learning Rate & Training Budget & Batch Size \\
\midrule
DPO    & $1\times10^{-6}$ & 2 epochs            & 64 \\
CAT    & $2\times10^{-4}$ & 5 epochs       & 64 \\
NPO    & $1\times10^{-5}$ & 10 epochs      & 32 \\
RMU    & $1\times10^{-6}$ & 10 epochs    & 32 \\
Unbias & $2\times10^{-5}$ & early stopping & 32 \\
TV     & $8\times10^{-6}$ & 20 epochs      & 32 \\
\bottomrule
\end{tabular}

\endgroup
\end{table}


\begin{table}[t]
\centering
\small
\setlength{\tabcolsep}{10pt}
\renewcommand{\arraystretch}{1.08}

\caption{
Baseline risk scores at $M_0$.
Larger values indicate higher risk.
}
\label{tab:raw-baseline}

\begin{tabular}{@{}llccc@{}}
\toprule
Family & Size & Safety (S) & Privacy (P) & Fairness (F) \\
\midrule

\multirow{2}{*}{Llama}
& S & 0.6535 & 0.7043 & 0.3116 \\
& L & 0.7046 & 0.9893 & 0.3419 \\

\addlinespace[2pt]

\multirow{2}{*}{Gemma}
& S & 0.5685 & 0.9841 & 0.1649 \\
& L & 0.6592 & 0.9963 & 0.1834 \\

\addlinespace[2pt]

\multirow{2}{*}{Qwen}
& S & 0.5718 & 0.1866 & 0.2602 \\
& L & 0.6807 & 0.8367 & 0.2935 \\

\bottomrule
\end{tabular}
\end{table}

\section{Risk Scores}
\label{app:risk-scores}

Table~\ref{tab:raw-baseline} reports the baseline risk scores at
$M_0$, and Tables~\ref{tab:raw-llama}--\ref{tab:raw-qwen} report
the checkpoint-level scores for all ordered defense compositions.
Here, S, P, and F denote safety, privacy, and fairness, respectively.
All scores are oriented so that larger values indicate higher risk.

For an ordered composition $D_1 \rightarrow D_2$, the target
dimension of each defense is shown in parentheses after its name.
Under each model, the \emph{Earlier} columns report
$S_{R_1}(M_1)$ and $S_{R_1}(M_2)$, while the
\emph{Subsequent} columns report
$S_{R_2}(M_1)$ and $S_{R_2}(M_2)$.
We retain $S_{R_2}(M_1)$ because the gain of the subsequent defense
is defined from the checkpoint at which that defense is applied,
$G_2=S_{R_2}(M_1)-S_{R_2}(M_2)$.
Comparing $M_2$ only with $M_0$ would instead measure the net change
from the initial model and would not isolate the effect of $D_2$.

Because all compositions for a given model share the same $M_0$,
the baseline scores are reported once in Table~\ref{tab:raw-baseline}
rather than repeated in every composition row.
All displayed scores are rounded to four decimal places for
readability.
All defense gains, regressions, RRR values, and interaction
classifications are computed from the corresponding unrounded
checkpoint-level scores.


\begin{table}[t]
\centering
\caption{
Checkpoint-level risk scores for the Llama family.
For each model, \emph{Earlier} denotes the target of $D_1$ and
\emph{Subsequent} denotes the target of $D_2$.
}
\label{tab:raw-llama}
\label{tab:raw-llamaS}
\label{tab:raw-llamaL}

\begingroup
\scriptsize
\setlength{\tabcolsep}{2.7pt}
\renewcommand{\arraystretch}{1.07}

\begin{tabular*}{\linewidth}{
@{\extracolsep{\fill}}
ll
rrrr
rrrr
@{}
}
\toprule

\multirow{3}{*}{\makecell[c]{Primary Defense\\($D_1$)}}
&
\multirow{3}{*}{\makecell[c]{Subsequent Defense\\($D_2$)}}
&
\multicolumn{4}{c}{Llama-S}
&
\multicolumn{4}{c}{Llama-L}
\\

\cmidrule(lr){3-6}
\cmidrule(lr){7-10}

&
&
\multicolumn{2}{c}{Earlier}
&
\multicolumn{2}{c}{Subsequent}
&
\multicolumn{2}{c}{Earlier}
&
\multicolumn{2}{c}{Subsequent}
\\

\cmidrule(lr){3-4}
\cmidrule(lr){5-6}
\cmidrule(lr){7-8}
\cmidrule(lr){9-10}

&
&
$M_1$ & $M_2$
&
$M_1$ & $M_2$
&
$M_1$ & $M_2$
&
$M_1$ & $M_2$
\\

\midrule

\multirow{4}{*}{\textbf{DPO (S)}}
& NPO (P)
& 0.4670 & 0.5095 & 0.7012 & 0.0562
& 0.5289 & 0.4092 & 0.9881 & 0.0329 \\

& RMU (P)
& 0.4670 & 0.5248 & 0.7012 & 0.0657
& 0.5289 & 0.4736 & 0.9881 & 0.2484 \\

& Unbias (F)
& 0.4670 & 0.4175 & 0.3037 & 0.0054
& 0.5289 & 0.3589 & 0.3421 & 0.0212 \\

& TV (F)
& 0.4670 & 0.3787 & 0.3037 & 0.2586
& 0.5289 & 0.1988 & 0.3421 & 0.1003 \\

\cmidrule(lr){1-10}

\multirow{4}{*}{\textbf{CAT (S)}}
& NPO (P)
& 0.2954 & 0.2591 & 0.1355 & 0.0993
& 0.2731 & 0.1922 & 0.2180 & 0.1252 \\

& RMU (P)
& 0.2954 & 0.2335 & 0.1355 & 0.0731
& 0.2731 & 0.2682 & 0.2180 & 0.0542 \\

& Unbias (F)
& 0.2954 & 0.2525 & 0.2890 & 0.0360
& 0.2731 & 0.2277 & 0.3642 & 0.0975 \\

& TV (F)
& 0.2954 & 0.2616 & 0.2890 & 0.2036
& 0.2731 & 0.2871 & 0.3642 & 0.0540 \\

\cmidrule(lr){1-10}

\multirow{4}{*}{\textbf{NPO (P)}}
& DPO (S)
& 0.0748 & 0.0729 & 0.5421 & 0.2591
& 0.4860 & 0.2529 & 0.6898 & 0.6724 \\

& CAT (S)
& 0.0748 & 0.1030 & 0.5421 & 0.2979
& 0.4860 & 0.1556 & 0.6898 & 0.2632 \\

& Unbias (F)
& 0.0748 & 0.0734 & 0.2618 & 0.0500
& 0.4860 & 0.2023 & 0.3600 & 0.0189 \\

& TV (F)
& 0.0748 & 0.1533 & 0.2618 & 0.1595
& 0.4860 & 0.1644 & 0.3600 & 0.0024 \\

\cmidrule(lr){1-10}

\multirow{4}{*}{\textbf{RMU (P)}}
& DPO (S)
& 0.0722 & 0.0358 & 0.6386 & 0.5553
& 0.2418 & 0.0466 & 0.6840 & 0.6139 \\

& CAT (S)
& 0.0722 & 0.1058 & 0.6386 & 0.3894
& 0.2418 & 0.0546 & 0.6840 & 0.3391 \\

& Unbias (F)
& 0.0722 & 0.0349 & 0.2966 & 0.0897
& 0.2418 & 0.0350 & 0.3402 & 0.0190 \\

& TV (F)
& 0.0722 & 0.0361 & 0.2966 & 0.2034
& 0.2418 & 0.0512 & 0.3402 & 0.0112 \\

\cmidrule(lr){1-10}

\multirow{4}{*}{\textbf{Unbias (F)}}
& DPO (S)
& 0.1128 & 0.0563 & 0.5528 & 0.4208
& 0.0927 & 0.0476 & 0.6790 & 0.5272 \\

& CAT (S)
& 0.1128 & 0.2321 & 0.5528 & 0.3234
& 0.0927 & 0.2730 & 0.6790 & 0.3375 \\

& NPO (P)
& 0.1128 & 0.0331 & 0.3015 & 0.0382
& 0.0927 & 0.0373 & 0.9487 & 0.0346 \\

& RMU (P)
& 0.1128 & 0.1118 & 0.3015 & 0.0468
& 0.0927 & 0.0548 & 0.9487 & 0.0325 \\

\cmidrule(lr){1-10}

\multirow{4}{*}{\textbf{TV (F)}}
& DPO (S)
& 0.1699 & 0.1828 & 0.5710 & 0.4637
& 0.0644 & 0.0641 & 0.5941 & 0.4455 \\

& CAT (S)
& 0.1699 & 0.2270 & 0.5710 & 0.3746
& 0.0644 & 0.1605 & 0.5941 & 0.2781 \\

& NPO (P)
& 0.1699 & 0.1714 & 0.1262 & 0.0356
& 0.0644 & 0.0174 & 0.3262 & 0.0435 \\

& RMU (P)
& 0.1699 & 0.1719 & 0.1262 & 0.0375
& 0.0644 & 0.0272 & 0.3262 & 0.0589 \\

\bottomrule
\end{tabular*}

\endgroup
\end{table}


\begin{table}[t]
\centering
\caption{
Checkpoint-level risk scores for the Gemma family.
For each model, \emph{Earlier} denotes the target of $D_1$ and
\emph{Subsequent} denotes the target of $D_2$.
}
\label{tab:raw-gemma}
\label{tab:raw-gemmaS}
\label{tab:raw-gemmaL}

\begingroup
\scriptsize
\setlength{\tabcolsep}{2.7pt}
\renewcommand{\arraystretch}{1.07}

\begin{tabular*}{\linewidth}{
@{\extracolsep{\fill}}
ll
rrrr
rrrr
@{}
}
\toprule

\multirow{3}{*}{\makecell[c]{Primary Defense\\($D_1$)}}
&
\multirow{3}{*}{\makecell[c]{Subsequent Defense\\($D_2$)}}
&
\multicolumn{4}{c}{Gemma-S}
&
\multicolumn{4}{c}{Gemma-L}
\\

\cmidrule(lr){3-6}
\cmidrule(lr){7-10}

&
&
\multicolumn{2}{c}{Earlier}
&
\multicolumn{2}{c}{Subsequent}
&
\multicolumn{2}{c}{Earlier}
&
\multicolumn{2}{c}{Subsequent}
\\

\cmidrule(lr){3-4}
\cmidrule(lr){5-6}
\cmidrule(lr){7-8}
\cmidrule(lr){9-10}

&
&
$M_1$ & $M_2$
&
$M_1$ & $M_2$
&
$M_1$ & $M_2$
&
$M_1$ & $M_2$
\\

\midrule

\multirow{4}{*}{\textbf{DPO (S)}}
& NPO (P)
& 0.4051 & 0.5190 & 0.9761 & 0.0512
& 0.2673 & 0.1444 & 0.9919 & 0.0306 \\

& RMU (P)
& 0.4051 & 0.4266 & 0.9761 & 0.1064
& 0.2673 & 0.2706 & 0.9919 & 0.2220 \\

& Unbias (F)
& 0.4051 & 0.5355 & 0.2080 & 0.0287
& 0.2673 & 0.2880 & 0.1039 & 0.0406 \\

& TV (F)
& 0.4051 & 0.4876 & 0.2080 & 0.0579
& 0.2673 & 0.1139 & 0.1039 & 0.0013 \\

\cmidrule(lr){1-10}

\multirow{4}{*}{\textbf{CAT (S)}}
& NPO (P)
& 0.3127 & 0.3358 & 0.1477 & 0.0876
& 0.2913 & 0.3729 & 0.1427 & 0.0306 \\

& RMU (P)
& 0.3127 & 0.2880 & 0.1477 & 0.0624
& 0.2913 & 0.3053 & 0.1427 & 0.0525 \\

& Unbias (F)
& 0.3127 & 0.2203 & 0.1601 & 0.0459
& 0.2913 & 0.4417 & 0.1472 & 0.0480 \\

& TV (F)
& 0.3127 & 0.2451 & 0.1601 & 0.0259
& 0.2913 & 0.1634 & 0.1472 & 0.0395 \\

\cmidrule(lr){1-10}

\multirow{4}{*}{\textbf{NPO (P)}}
& DPO (S)
& 0.0306 & 0.0306 & 0.8078 & 0.6081
& 0.0306 & 0.0306 & 0.5215 & 0.3837 \\

& CAT (S)
& 0.0306 & 0.0657 & 0.8078 & 0.2401
& 0.0306 & 0.0458 & 0.5215 & 0.3507 \\

& Unbias (F)
& 0.0306 & 0.0306 & 0.1624 & 0.0867
& 0.0306 & 0.0306 & 0.1587 & 0.0314 \\

& TV (F)
& 0.0306 & 0.0306 & 0.1624 & 0.1616
& 0.0306 & 0.0306 & 0.1587 & 0.0224 \\

\cmidrule(lr){1-10}

\multirow{4}{*}{\textbf{RMU (P)}}
& DPO (S)
& 0.1695 & 0.1875 & 0.6238 & 0.4026
& 0.1890 & 0.0611 & 0.6823 & 0.2756 \\

& CAT (S)
& 0.1695 & 0.1245 & 0.6238 & 0.1848
& 0.1890 & 0.0431 & 0.6823 & 0.4010 \\

& Unbias (F)
& 0.1695 & 0.0310 & 0.1049 & 0.0133
& 0.1890 & 0.0357 & 0.1243 & 0.0599 \\

& TV (F)
& 0.1695 & 0.0306 & 0.1049 & 0.0797
& 0.1890 & 0.0306 & 0.1243 & 0.0518 \\

\cmidrule(lr){1-10}

\multirow{4}{*}{\textbf{Unbias (F)}}
& DPO (S)
& 0.0142 & 0.0750 & 0.7343 & 0.4563
& 0.0502 & 0.0103 & 0.6370 & 0.2797 \\

& CAT (S)
& 0.0142 & 0.1102 & 0.7343 & 0.2434
& 0.0502 & 0.0903 & 0.6370 & 0.5173 \\

& NPO (P)
& 0.0142 & 0.0404 & 0.4603 & 0.0306
& 0.0502 & 0.0081 & 0.9799 & 0.0306 \\

& RMU (P)
& 0.0142 & 0.0374 & 0.4603 & 0.0315
& 0.0502 & 0.0789 & 0.9799 & 0.1528 \\

\cmidrule(lr){1-10}

\multirow{4}{*}{\textbf{TV (F)}}
& DPO (S)
& 0.0841 & 0.0487 & 0.6328 & 0.3705
& 0.0291 & 0.0535 & 0.5347 & 0.1667 \\

& CAT (S)
& 0.0841 & 0.1306 & 0.6328 & 0.2781
& 0.0291 & 0.0581 & 0.5347 & 0.4975 \\

& NPO (P)
& 0.0841 & 0.1092 & 0.3052 & 0.0306
& 0.0291 & 0.0290 & 0.2305 & 0.0306 \\

& RMU (P)
& 0.0841 & 0.0124 & 0.3052 & 0.0306
& 0.0291 & 0.0891 & 0.2305 & 0.0306 \\

\bottomrule
\end{tabular*}

\endgroup
\end{table}


\begin{table}[t]
\centering
\caption{
Checkpoint-level risk scores for the Qwen family.
For each model, \emph{Earlier} denotes the target of $D_1$ and
\emph{Subsequent} denotes the target of $D_2$.
}
\label{tab:raw-qwen}
\label{tab:raw-qwenS}
\label{tab:raw-qwenL}

\begingroup
\scriptsize
\setlength{\tabcolsep}{2.7pt}
\renewcommand{\arraystretch}{1.07}

\begin{tabular*}{\linewidth}{
@{\extracolsep{\fill}}
ll
rrrr
rrrr
@{}
}
\toprule

\multirow{3}{*}{\makecell[c]{Primary Defense\\($D_1$)}}
&
\multirow{3}{*}{\makecell[c]{Subsequent Defense\\($D_2$)}}
&
\multicolumn{4}{c}{Qwen-S}
&
\multicolumn{4}{c}{Qwen-L}
\\

\cmidrule(lr){3-6}
\cmidrule(lr){7-10}

&
&
\multicolumn{2}{c}{Earlier}
&
\multicolumn{2}{c}{Subsequent}
&
\multicolumn{2}{c}{Earlier}
&
\multicolumn{2}{c}{Subsequent}
\\

\cmidrule(lr){3-4}
\cmidrule(lr){5-6}
\cmidrule(lr){7-8}
\cmidrule(lr){9-10}

&
&
$M_1$ & $M_2$
&
$M_1$ & $M_2$
&
$M_1$ & $M_2$
&
$M_1$ & $M_2$
\\

\midrule

\multirow{4}{*}{\textbf{DPO (S)}}
& NPO (P)
& 0.4472 & 0.3878 & 0.1836 & 0.0770
& 0.5528 & 0.5446 & 0.8171 & 0.8153 \\

& RMU (P)
& 0.4472 & 0.4307 & 0.1836 & 0.1243
& 0.5528 & 0.5264 & 0.8171 & 0.8049 \\

& Unbias (F)
& 0.4472 & 0.4290 & 0.2605 & 0.0007
& 0.5528 & 0.4282 & 0.2936 & 0.0481 \\

& TV (F)
& 0.4472 & 0.2657 & 0.2605 & 0.1532
& 0.5528 & 0.1865 & 0.2936 & 0.1911 \\

\cmidrule(lr){1-10}

\multirow{4}{*}{\textbf{CAT (S)}}
& NPO (P)
& 0.1436 & 0.1799 & 0.0834 & 0.0300
& 0.0058 & 0.0033 & 0.0927 & 0.0902 \\

& RMU (P)
& 0.1436 & 0.1139 & 0.0834 & 0.0727
& 0.0058 & 0.0041 & 0.0927 & 0.0926 \\

& Unbias (F)
& 0.1436 & 0.1601 & 0.2610 & 0.0341
& 0.0058 & 0.0074 & 0.3240 & 0.0655 \\

& TV (F)
& 0.1436 & 0.2723 & 0.2610 & 0.1245
& 0.0058 & 0.2871 & 0.3240 & 0.0547 \\

\cmidrule(lr){1-10}

\multirow{4}{*}{\textbf{NPO (P)}}
& DPO (S)
& 0.0415 & 0.0302 & 0.5635 & 0.4744
& 0.0309 & 0.0312 & 0.6411 & 0.4373 \\

& CAT (S)
& 0.0415 & 0.0402 & 0.5635 & 0.1064
& 0.0309 & 0.0673 & 0.6411 & 0.0033 \\

& Unbias (F)
& 0.0415 & 0.0302 & 0.2413 & 0.0354
& 0.0309 & 0.0304 & 0.2952 & 0.0447 \\

& TV (F)
& 0.0415 & 0.0300 & 0.2413 & 0.1299
& 0.0309 & 0.0300 & 0.2952 & 0.1711 \\

\cmidrule(lr){1-10}

\multirow{4}{*}{\textbf{RMU (P)}}
& DPO (S)
& 0.1028 & 0.1042 & 0.5792 & 0.4505
& 0.1774 & 0.1314 & 0.7104 & 0.5182 \\

& CAT (S)
& 0.1028 & 0.0730 & 0.5792 & 0.1205
& 0.1774 & 0.0916 & 0.7104 & 0.0017 \\

& Unbias (F)
& 0.1028 & 0.0710 & 0.2515 & 0.0250
& 0.1774 & 0.1087 & 0.3058 & 0.0435 \\

& TV (F)
& 0.1028 & 0.0300 & 0.2515 & 0.1188
& 0.1774 & 0.0300 & 0.3058 & 0.1979 \\

\cmidrule(lr){1-10}

\multirow{4}{*}{\textbf{Unbias (F)}}
& DPO (S)
& 0.0003 & 0.0129 & 0.6147 & 0.4109
& 0.0323 & 0.0118 & 0.6939 & 0.4579 \\

& CAT (S)
& 0.0003 & 0.1717 & 0.6147 & 0.1304
& 0.0323 & 0.1697 & 0.6939 & 0.0066 \\

& NPO (P)
& 0.0003 & 0.0035 & 0.1307 & 0.0467
& 0.0323 & 0.0847 & 0.6076 & 0.0763 \\

& RMU (P)
& 0.0003 & 0.0073 & 0.1307 & 0.0668
& 0.0323 & 0.0510 & 0.6076 & 0.0378 \\

\cmidrule(lr){1-10}

\multirow{4}{*}{\textbf{TV (F)}}
& DPO (S)
& 0.1162 & 0.1188 & 0.4893 & 0.3325
& 0.1017 & 0.0996 & 0.4761 & 0.3135 \\

& CAT (S)
& 0.1162 & 0.2039 & 0.4893 & 0.1081
& 0.1017 & 0.1436 & 0.4761 & 0.0000 \\

& NPO (P)
& 0.1162 & 0.1204 & 0.1300 & 0.0300
& 0.1017 & 0.1129 & 0.1300 & 0.0300 \\

& RMU (P)
& 0.1162 & 0.1031 & 0.1300 & 0.0300
& 0.1017 & 0.0068 & 0.1300 & 0.0300 \\

\bottomrule
\end{tabular*}

\endgroup
\end{table}

\section{General Utility Evaluation}
\label{app:mmlu}

To examine whether the observed defense interactions could be explained by broad capability degradation, we additionally evaluate MMLU accuracy for all sequential defense compositions. As shown in \autoref{tab:mmlu_all_scenarios}, most compositions remain within a few percentage points of their corresponding undefended baselines. This suggests that the risk regressions reported in \autoref{subsec:measurement-results} are not generally attributable to an overall collapse in model capability. Some compositions, particularly those involving Task Vector (TV) editing, exhibit somewhat larger decreases in MMLU accuracy, indicating that general-utility degradation can occur in particular settings and should be considered separately from sequential defense compatibility.

\begin{table*}[t]
\centering
\caption{
MMLU accuracy (\%) across all sequential defense compositions, reported as
mean $\pm$ std. Most compositions remain close to the corresponding
undefended baseline, suggesting no general capability collapse.
}
\label{tab:mmlu_all_scenarios}

\begingroup
\small
\setlength{\tabcolsep}{4pt}
\renewcommand{\arraystretch}{1.05}

\resizebox{\linewidth}{!}{%
\begin{tabular}{@{}llcccccc@{}}
\toprule
\multirow{2}{*}{\textbf{Primary Defense ($D_1$)}} 
  & \multirow{2}{*}{\textbf{Subsequent Defense ($D_2$)}} 
  & \multicolumn{2}{c}{\textbf{Llama}} 
  & \multicolumn{2}{c}{\textbf{Gemma}} 
  & \multicolumn{2}{c}{\textbf{Qwen}} \\
\cmidrule(lr){3-4}
\cmidrule(lr){5-6}
\cmidrule(lr){7-8}
& & \textbf{-S} & \textbf{-L}
  & \textbf{-S} & \textbf{-L}
  & \textbf{-S} & \textbf{-L} \\
\midrule

\textbf{No Defense} & --
  & $46.03{\scriptstyle\pm0.41}$ & $66.10{\scriptstyle\pm0.38}$
  & $56.44{\scriptstyle\pm0.40}$ & $51.70{\scriptstyle\pm0.39}$
  & $59.75{\scriptstyle\pm0.39}$ & $72.33{\scriptstyle\pm0.36}$ \\
\midrule

\multicolumn{8}{c}{\textit{Fairness-First}} \\
\midrule

\multirow{4}{*}{\shortstack[l]{\textbf{Unbias}\\(Fairness Stage)}}
  & + RMU (Privacy)
    & $44.05{\scriptstyle\pm0.41}$ & $65.18{\scriptstyle\pm0.39}$
    & $54.45{\scriptstyle\pm0.40}$ & $51.24{\scriptstyle\pm0.40}$
    & $57.63{\scriptstyle\pm0.40}$ & $71.73{\scriptstyle\pm0.38}$ \\
& + NPO (Privacy)
    & $45.39{\scriptstyle\pm0.41}$ & $65.35{\scriptstyle\pm0.38}$
    & $56.02{\scriptstyle\pm0.40}$ & $51.70{\scriptstyle\pm0.39}$
    & $59.69{\scriptstyle\pm0.39}$ & $70.89{\scriptstyle\pm0.40}$ \\
\cmidrule(l){2-8}
& + DPO (Safety)
    & $45.85{\scriptstyle\pm0.41}$ & $65.63{\scriptstyle\pm0.38}$
    & $56.27{\scriptstyle\pm0.40}$ & $50.75{\scriptstyle\pm0.39}$
    & $59.59{\scriptstyle\pm0.39}$ & $72.16{\scriptstyle\pm0.36}$ \\
& + CAT (Safety)
    & $43.71{\scriptstyle\pm0.41}$ & $64.91{\scriptstyle\pm0.39}$
    & $53.73{\scriptstyle\pm0.40}$ & $50.13{\scriptstyle\pm0.38}$
    & $59.28{\scriptstyle\pm0.39}$ & $71.19{\scriptstyle\pm0.36}$ \\
\midrule

\multirow{4}{*}{\shortstack[l]{\textbf{Task Vector (TV)}\\(Fairness Stage)}}
  & + RMU (Privacy)
    & $43.95{\scriptstyle\pm0.40}$ & $63.86{\scriptstyle\pm0.40}$
    & $54.78{\scriptstyle\pm0.41}$ & $50.31{\scriptstyle\pm0.40}$
    & $58.44{\scriptstyle\pm0.41}$ & $70.40{\scriptstyle\pm0.37}$ \\
& + NPO (Privacy)
    & $44.20{\scriptstyle\pm0.41}$ & $64.66{\scriptstyle\pm0.41}$
    & $54.94{\scriptstyle\pm0.41}$ & $50.52{\scriptstyle\pm0.39}$
    & $57.55{\scriptstyle\pm0.40}$ & $71.58{\scriptstyle\pm0.41}$ \\
\cmidrule(l){2-8}
& + DPO (Safety)
    & $44.12{\scriptstyle\pm0.41}$ & $64.64{\scriptstyle\pm0.41}$
    & $54.46{\scriptstyle\pm0.41}$ & $51.04{\scriptstyle\pm0.39}$
    & $56.92{\scriptstyle\pm0.41}$ & $71.40{\scriptstyle\pm0.41}$ \\
& + CAT (Safety)
    & $43.48{\scriptstyle\pm0.41}$ & $63.81{\scriptstyle\pm0.39}$
    & $53.78{\scriptstyle\pm0.40}$ & $49.74{\scriptstyle\pm0.40}$
    & $58.63{\scriptstyle\pm0.39}$ & $70.38{\scriptstyle\pm0.37}$ \\
\midrule

\multicolumn{8}{c}{\textit{Privacy-First}} \\
\midrule

\multirow{4}{*}{\shortstack[l]{\textbf{NPO}\\(Privacy Stage)}}
  & + Unbias (Fairness)
    & $45.82{\scriptstyle\pm0.41}$ & $65.03{\scriptstyle\pm0.38}$
    & $56.65{\scriptstyle\pm0.40}$ & $51.31{\scriptstyle\pm0.40}$
    & $58.42{\scriptstyle\pm0.40}$ & $72.58{\scriptstyle\pm0.36}$ \\
& + TV (Fairness)
    & $44.13{\scriptstyle\pm0.41}$ & $63.86{\scriptstyle\pm0.38}$
    & $54.43{\scriptstyle\pm0.41}$ & $50.01{\scriptstyle\pm0.39}$
    & $58.35{\scriptstyle\pm0.40}$ & $71.09{\scriptstyle\pm0.39}$ \\
\cmidrule(l){2-8}
& + DPO (Safety)
    & $44.50{\scriptstyle\pm0.41}$ & $65.40{\scriptstyle\pm0.38}$
    & $56.50{\scriptstyle\pm0.40}$ & $51.84{\scriptstyle\pm0.41}$
    & $59.76{\scriptstyle\pm0.39}$ & $72.55{\scriptstyle\pm0.36}$ \\
& + CAT (Safety)
    & $43.78{\scriptstyle\pm0.41}$ & $64.27{\scriptstyle\pm0.39}$
    & $53.75{\scriptstyle\pm0.40}$ & $51.02{\scriptstyle\pm0.40}$
    & $59.31{\scriptstyle\pm0.39}$ & $71.62{\scriptstyle\pm0.36}$ \\
\midrule

\multirow{4}{*}{\shortstack[l]{\textbf{RMU}\\(Privacy Stage)}}
  & + Unbias (Fairness)
    & $45.21{\scriptstyle\pm0.41}$ & $64.25{\scriptstyle\pm0.38}$
    & $54.99{\scriptstyle\pm0.40}$ & $51.11{\scriptstyle\pm0.38}$
    & $58.56{\scriptstyle\pm0.40}$ & $69.62{\scriptstyle\pm0.37}$ \\
& + TV (Fairness)
    & $44.39{\scriptstyle\pm0.40}$ & $63.57{\scriptstyle\pm0.39}$
    & $53.92{\scriptstyle\pm0.41}$ & $50.88{\scriptstyle\pm0.40}$
    & $57.49{\scriptstyle\pm0.41}$ & $70.75{\scriptstyle\pm0.41}$ \\
\cmidrule(l){2-8}
& + DPO (Safety)
    & $44.00{\scriptstyle\pm0.41}$ & $64.68{\scriptstyle\pm0.38}$
    & $55.96{\scriptstyle\pm0.40}$ & $50.92{\scriptstyle\pm0.41}$
    & $59.00{\scriptstyle\pm0.40}$ & $70.60{\scriptstyle\pm0.37}$ \\
& + CAT (Safety)
    & $43.23{\scriptstyle\pm0.41}$ & $64.41{\scriptstyle\pm0.39}$
    & $53.80{\scriptstyle\pm0.40}$ & $50.08{\scriptstyle\pm0.38}$
    & $58.70{\scriptstyle\pm0.39}$ & $70.87{\scriptstyle\pm0.36}$ \\
\midrule

\multicolumn{8}{c}{\textit{Safety-First}} \\
\midrule

\multirow{4}{*}{\shortstack[l]{\textbf{DPO}\\(Safety Stage)}}
  & + Unbias (Fairness)
    & $45.41{\scriptstyle\pm0.38}$ & $64.58{\scriptstyle\pm0.38}$
    & $53.42{\scriptstyle\pm0.41}$ & $51.22{\scriptstyle\pm0.40}$
    & $59.46{\scriptstyle\pm0.39}$ & $72.37{\scriptstyle\pm0.36}$ \\
& + TV (Fairness)
    & $44.07{\scriptstyle\pm0.41}$ & $63.01{\scriptstyle\pm0.40}$
    & $53.95{\scriptstyle\pm0.41}$ & $50.26{\scriptstyle\pm0.41}$
    & $57.34{\scriptstyle\pm0.41}$ & $71.01{\scriptstyle\pm0.40}$ \\
\cmidrule(l){2-8}
& + NPO (Privacy)
    & $44.64{\scriptstyle\pm0.41}$ & $64.73{\scriptstyle\pm0.38}$
    & $56.05{\scriptstyle\pm0.40}$ & $51.38{\scriptstyle\pm0.40}$
    & $59.77{\scriptstyle\pm0.39}$ & $72.15{\scriptstyle\pm0.37}$ \\
& + RMU (Privacy)
    & $43.90{\scriptstyle\pm0.41}$ & $63.56{\scriptstyle\pm0.39}$
    & $55.08{\scriptstyle\pm0.40}$ & $51.34{\scriptstyle\pm0.39}$
    & $57.71{\scriptstyle\pm0.40}$ & $71.94{\scriptstyle\pm0.37}$ \\
\midrule

\multirow{4}{*}{\shortstack[l]{\textbf{CAT}\\(Safety Stage)}}
  & + Unbias (Fairness)
    & $44.53{\scriptstyle\pm0.41}$ & $65.79{\scriptstyle\pm0.39}$
    & $54.26{\scriptstyle\pm0.40}$ & $51.00{\scriptstyle\pm0.40}$
    & $59.30{\scriptstyle\pm0.39}$ & $70.99{\scriptstyle\pm0.36}$ \\
& + TV (Fairness)
    & $43.81{\scriptstyle\pm0.37}$ & $64.65{\scriptstyle\pm0.36}$
    & $53.65{\scriptstyle\pm0.36}$ & $49.59{\scriptstyle\pm0.41}$
    & $57.45{\scriptstyle\pm0.41}$ & $69.89{\scriptstyle\pm0.37}$ \\
\cmidrule(l){2-8}
& + NPO (Privacy)
    & $44.80{\scriptstyle\pm0.41}$ & $64.76{\scriptstyle\pm0.39}$
    & $54.33{\scriptstyle\pm0.40}$ & $51.61{\scriptstyle\pm0.40}$
    & $59.14{\scriptstyle\pm0.39}$ & $71.90{\scriptstyle\pm0.37}$ \\
& + RMU (Privacy)
    & $43.89{\scriptstyle\pm0.41}$ & $63.72{\scriptstyle\pm0.40}$
    & $54.82{\scriptstyle\pm0.40}$ & $51.44{\scriptstyle\pm0.41}$
    & $58.21{\scriptstyle\pm0.40}$ & $71.80{\scriptstyle\pm0.37}$ \\
\bottomrule
\end{tabular}%
}

\endgroup
\end{table*}

\end{document}